# Magnetron Sputtered NbN Films with Nb Interlayer

Kulwant Singh[#], A.C. Bidaye, A. K. Suri*

SES, MPD, *Materials Group,

Bhabha Atomic Research Centre, Trombay, Mumbai – 400085, India

[#]singhkw@barc.gov.in; Ph : +91-22-25595378; Fax : +91-22-25505151

## Abstract

NbN films were deposited on SS substrates by reactive DC magnetron sputtering at various $N_2$ flow rates and substrate biasing. Effect of $N_2$ flow rate and substrate biasing has been studied on deposition rate, surface hardness, crystal-structure and adhesion values. Process parameters were optimized for deposition of NbN coatings. NbN coatings were then deposited on to MS substrates as such and with Nb interlayer deposited by magnetron sputtering. The thickness of interlayer was 2 µm. The duplex coating has been studied for the improvement with respect to surface hardness by Knoop micro indentation and corrosion performance by potentiodynamic polarization technique. Open circuit potentials were also measured.

## 1. Introduction

Magnetron Sputtering, a physical vapour deposition technique, is widely used for deposition of compound coatings [1,2,3,4,5]. Various types of coatings involving binary, ternary, multi component compound coatings, multilayers, duplex coatings, nano-crystalline coatings etc are increasingly being studied; as these find variety of applications due to their exotic properties. Niobium nitride (NbN) films in the initial years were investigated more because of their superconducting properties rather than their mechanical properties. The research for synthesis of NbN films was directed to increase their superconducting transition temperatures [6,7,8]. However, NbN films are attractive in wear applications too because of their good thermal expansion match with widely used tool steels. Good mechanical properties coupled with chemical inertness, wear resistance, high melting point, temperature stability and high electrical conductivity make NbN films a suitable material for protective coating [9], field emission cathode [10] and diffusion barrier in microelectronic devices [11]. The deposition of NbN films have been carried out by various techniques including reactive magnetron sputtering [12,13,14,15], ion beam assisted deposition [16,17], pulsed laser deposition [18] and cathodic arc deposition [19, 20, 21, 22]. After the introduction of superlattice coatings, NbN was used as one layer component. Superlattice coatings such as TiN/NbN [23,24,25], TaN/NbN [26] and CrN/NbN [27,28] have been investigated for use as hard, wear resistant and corrosion protective coatings. All of these superlattice films possess super-hardness effects, which exhibit an anomalous increase in hardness and wear resistance.

NbN coatings with interlayer, however, have not been concentrated upon. Duplex coatings involving TiN as top coat deposited by sputtering with interlayer of chromium (Cr) and nickel (Ni) deposited by electroplating and electroless nickel (EN) deposited by electroless techniques have been found to improve hardness, corrosion resistance and other properties when extended on to mild steel (MS) substrates [29,30,31,32]. In the present study, NbN coating deposited by reactive magnetron sputtering has been extended on to MS substrate with interlayer of niobium (Nb) deposited by sputtering to explore the benefits on MS substrates. For this NbN coatings were first deposited on to stainless steel (SS) substrates; coatings were studied for their structure, thickness, hardness and adhesion; process parameters optimized and then NbN coatings as such and with Nb interlayer were deposited on to MS substrates and studied for the improvement with respect to surface hardness by Knoop micro indentation and corrosion performance by potentiodynamic polarization technique. Open circuit potentials were also measured.

## 2. Experimental Procedure

The NbN films were deposited using reactive DC magnetron sputtering on SS, MS and Nb coated MS substrates. A Nb (99.99% purity) metallic target, 160 mm diameter and 4 mm thick, was mechanically clamped to a planar sputter source mounted horizontally on the base of the chamber evacuated to a base pressure of $2\times10^{-6}$ mbar. The distance between the target and the substrate was set to be 60 mm. The sputtering pressure was kept at $5\times10^{-3}$ mbar by admitting a stream of mixed gas of Ar and $N_2$ into the chamber. Flow of Argon (Ar) gas was fixed at 20sccm while $N_2$ flow was varied from 0-14sccm. Substrate biasing was kept constant at –50V for coatings deposited at different nitrogen flow ratios. The power to the target was supplied through a stabilized d. c. power supply of 0-1000V (6 Amperes maximum). Substrate biasing was varied from 0 to -150V in a step of 25V (keeping the $N_2$/Ar flow ratio constant at 20%) by means of a stabilized d. c. power supply of variable voltage (0-300V) and current (0-500mA). The samples were polished, cleaned thoroughly and degreased in alkaline solution prior to deposition. Proprietary mixtures were used for the purpose. The flow sheet used for the cleaning of the substrate samples before deposition is presented in the figure 1. All the cleaning steps were performed ultrasonically for the specified durations. The mild steel samples were deposited with Nb as interlayer before deposition of NbN coatings. The deposition parameters for Nb and NbN coatings were similar (Table1) except that no $N_2$ was introduced for Nb coating. Thickness of Nb interlayer was kept at 2 µm.

Weight gains of the samples were recorded and thickness of the coatings was calculated using bulk density values. Actual coating thicknesses were studied by Calotest techniques using 52100 steel ball with micro abrasion by diamond particles. The phase structures of the films were investigated by X-ray diffraction (XRD) with CuKα radiation. Surface hardness was measured by a microhardness tester (Future Tech FM-7 model) using Knoop indenter at loads of 100, 50 and 25gf. Depths of indentations were also measured. Five readings were performed for each of the coated sample and the average values reported. NbN coatings on MS and SS were evaluated for their adhesion performance by scratch adhesion tester at different loading rates of 10N/min, 30N/min 50N/min and 80N/min. The scratch length was 3 mm. The scratch indenter used was a 200µm tip radius Rockwell type diamond indenter. Friction force and depth of indentation for all the scratched samples were recorded online. The scratch tracks were seen in the optical microscopy immediately after the tests to visualize the scratch patterns and pictures were taken at different loads. The starting load in each test was 1N while maximum load was varied from 10N to 60N. Tests were performed in a linearly progressive mode from 1N start load to a predefined maximum load.

Electrochemical evaluation of the coated samples was carried out using the standard potentiodynamic measurement technique with a computer controlled Santronic Electrochemical Analyzer. Tests were carried out using a three-electrode cell. One side coated samples were soldered (with indium) on the other side to a copper wire coated with enamel. The samples were masked by Shailmask 800 lacquer (proprietary) to get the 1cm$^2$ surface area exposed. All potentials were measured with respect to a saturated calomel electrode (SCE). The auxiliary or counter electrode was platinum. The anodic and cathodic electrochemical polarization curves of all the samples were obtained in nitrogen de-aerated 1N $H_2SO_4$ electrolyte at room temperature. Open circuit potentials (OCP) were measured in deaerated 1N $H_2SO_4$ solution for 2 hrs. Before potentiodynamic measurements, the samples were allowed to reach equilibrium potential ($E_{ocp}$). This potential was reached after 30-40 minutes and the electrochemical measurements were started when the potential did not change by more than 1mV/min. The solution was replaced after each sweep run. Polarization resistance was determined in the ±15mV domain of $E_{ocp}$ potential using the linear polarization method at a scan rate of 0.1mV/sec. For potentiodynamic studies a potential sweep range of –1.000V to +1.000V was applied with a scan rate of 0.5mV/sec. The plot of E measured against SCE v/s log I was plotted. The corrosion potential ($E_{corr}$) was determined from the intersection of Tafel slopes and the corrosion current density ($I_{corr}$) was calculated using the anodic and cathodic Tafel slopes ($\beta_a$ and $\beta_c$) and polarization resistance ($R_p$).

## 3. Results and Discussion

### 3.1 Thickness

Weight gain method was utilized to calculate the thickness of the coatings, taking the bulk density value from the literature. Actual coating thicknesses were found out by Calotest technique, where abrasion of the coated samples was performed by a rotating 52100 steel ball using diamond particles. A variation of 5-20% was found in the calculated and actual thickness values due to the density of the coatings being lower than the bulk values. In the deposition using biasing there is a continuous ion bombardment at the substrate, causing the reduced effective deposition rate, which in turn reduces the coating thickness and therefore more time was required to get the same coating thickness for NbN coatings deposited at higher bias voltages. Coating thickness of about 1.8μm ±10% was obtained.

### 3.2 Deposition Rate
#### 3.2.1 Effect of Nitrogen Flow

Deposition rates of Nb-N coatings have been plotted in figure 2 as a function of $N_2$/Ar flow ratio. The coatings were deposited without substrate biasing and at a constant substrate bias voltage of –50V. Deposition rate for Nb-N films varied substantially with the variation in nitrogen flow ratio. Deposition rate decreased from 20nm/min to 10nm/min (without biasing) and from 16nm/min to 8nm/min (with biasing at -50V) with the increase in nitrogen flow from 0-70%. Deposition rate reduced almost linearly in both the cases with every increase in $N_2$ flow. Further, it can be seen that the reduction in deposition rate follows three different linear paths in both the cases – with biasing or without biasing. The three linear decreases in deposition rates correspond to different transition phases of Nb to $Nb_2N$; $Nb_2N$ to cubic NbN and cubic NbN to *h* NbN as revealed by XRD.

### 3.2.2 Effect of Substrate Biasing

Under biasing there is a continuous ion bombardment at the substrate, which imparts energy and thus not only improves adhesion, but also coating density. However, deposition rate is reduced. Figure 3 shows the deposition rate plotted against the substrate bias voltage keeping the flow ratio of nitrogen constant ($F_{N2}/F_{Ar}$=20%). It can be observed from the figure that deposition rate decreased with the increase in substrate biasing (negative) voltage. The deposition rate decreased from 15.9nm/min to 6nm/min when the bias voltage (negative) was increased from 0 to 150 volts. The effect of substrate biasing was found to be more pronounced than the effect of nitrogen flow ratio on the deposition rate.

The decrease in deposition rate with the increase in substrate biasing is due to the resputtering effect at the substrate. With every increase in substrate biasing (negative voltage) increased resputtering at the substrate (with higher energy ions) takes place causing the entrapped gas atoms and bonded particles with lower energies to be detached from the substrate resulting in fall of deposition rates.

### 3.3 Hardness
### 3.3.1 Effect of Nitrogen Flow

Knoop microhardness values for Nb-N coated on SS, taken at a load of 25gf, have been plotted as a function of increasing $N_2$ flow ratio in figure 4 (substrate biasing was kept constant at –50V). Surface hardness was found to increase rapidly with the increase in $N_2$ flow. The surface hardness value reached a maximum of 2040 HK at a nitrogen flow ratio of 20% and then started decreasing slowly with the further increase in nitrogen flow. The hardness decrease of the coating was accompanied with the observed changes in the crystalline structure of coatings as revealed by XRD discussed in section 3.4. Surface hardness values (at 25gf) for Nb-N films on SS have been summarized in Table 2 along with the depths of indentation. It is seen from the depths of indentation data that the surface hardness values depict the composite hardness and not the true hardness of the films even at a load of 25gf. Since, the maximum hardness value obtained was at 20% $N_2$ flow, further coatings of NbN as top coat with Nb interlayer on MS were carried out at 20% $N_2$ flow.

### 3.3.2 Effect of Substrate Biasing

Figure 5 shows the surface hardness values for NbN coatings on SS plotted against the substrate bias voltage keeping the nitrogen flow ratio constant (at 20%). Hardness increases continuously with the increase in substrate biasing voltage. The increase in hardness is due to the increased ion bombardment at the substrate with every increase in substrate biasing that leads to the increased coating density. Hardness of the coatings increased consistently from 1692 HK for coatings deposited without biasing to 2346 HK for coatings deposited at -150V substrate biasing.

### 3.3.3 Duplex Coatings on MS

The surface hardness of MS substrate, Nb, NbN and duplex coating of NbN with Nb interlayer on MS are given in Table 3. Nb coating was found to show the surface hardness value of about 434 HK. The substrate effect is pronounced in case of NbN coating on mild steel. Surface hardness of NbN coatings increased from 487 to 1084 with the decrease in applied load from 100gf to 25gf. With the incorporation of Nb interlayer, the surface hardness of NbN coatings on MS was found to increase due to the load support provided by the interlayer.

### 3.4 X-ray Diffraction

X-ray diffraction patterns of Nb-N films deposited on SS under various $N_2$ flow ratios are shown in figure 6. Coatings deposited at 5% nitrogen flow ratio show hexagonal $\beta$ $Nb_2N$ as the major phase with (101) preferred orientation. With increase in nitrogen flow to 10% the major phase becomes cubic $\delta$ NbN with preferred orientation of (111). At 30% $N_2$ flow hexagonal $\delta$' NbN phase appears, though the major phase is still cubic $\delta$ NbN but now with preferred orientation of (200). With further increase in $N_2$ flow the hexagonal $\delta$' NbN phase increases and becomes major phase at 70% $N_2$ flow. In all the coatings substrate peaks were identified as the major peaks.

Figure 7 shows the X-ray diffraction pattern of Nb coating, and NbN coating with Nb interlayer on MS substrate. Nb (110) was found to be stronger than Nb (200) peak. The intensity of Nb peaks reduced when NbN top coat was given. Mild steel samples were found to show the peak of Fe (110) in the 2θ range performed. NbN (111) peak was found to be stronger than NbN (200) or NbN (220) peaks.

### 3.5 Scratch Adhesion Test

Friction force and depth of indentation for all the scratch tests were recorded online along with the indenter movement to confirm the critical loads for cracks, chipping, delamination, coating failure or other observations. Tests performed at different loading rates were observed to give almost similar results and variation in loading rate had little impact on the scratching behaviour. Several types of observations were revealed as the scratch progressed in a linearly increasing load mode, such as - top layer removal, pile-up on the sides, visibility of small cracks to long wide cracks within the coatings, pores, chipping, partial or complete delamination of the coating etc. Figure 8 shows the scratch pattern for NbN coatings on MS sample taken at a load of 9N showing chipping, top layer removal and cracks. Figure 9 shows the scratch pattern for NbN coating on MS taken at 28N load revealing nearly complete delamination. Figure 10 exhibits the scratch pattern for NbN coating on SS taken at the start of the scratch (1N load). Figures 11-13 bare the scratch patterns for NbN coating on SS taken at a loads of 12N, 22N and 32N respectively showing the segregation, cracks within coating, pores, chipping, pile-up etc to an increasing extent.

#### 3.5.1 Critical Loads

Two critical loads $Lc_1$ and $Lc_2$ have been defined for the failure of the coatings. $Lc_1$ the first critical load corresponds to initial cohesive failure of the coating such as appearance of first cracks within the coating. $Lc_2$ the second critical load corresponds to adhesive failure of the coating i.e. first observation of adhesive failure such as chipping, partial delamination, pores or some such phenomena, where substrate beneath coating gets exposed. $Lc_1$ and $Lc_2$ values for coatings on MS samples were observed to be between 6-8N and 9-12N loads respectively. For coatings deposited on SS substrates $Lc_1$ values varied from 8-15N and $Lc_2$ values varied from 12-25N.

#### 3.5.2 Coefficient of Friction

Coefficient of friction (μ) as observed in the scratch adhesion test varied with the scratch load. For coatings deposited on SS samples, the value was found to vary within a narrow range of 0.22-0.25 at 30N load irrespective of coatings deposited at different flow ratio of $N_2$. At 40N load the value increased to 0.30 and at 60N load, the value was observed to be 0.40. For MS samples, the µ value was 0.28 for 30N load, 0.35 for 40N load and 0.45 for 60N load. Table 4 lists the µ value at different loads for the two types of substrates.

### 3.5.3 Effect of Loading Rate

The effect of loading rates on $Lc_1$ and $Lc_2$ was found to have little impact. The $Lc_1$ value was found to shift from 7N at 10N/min to 7.5N at 30N/min and further to 8.2N at 50N/min. Similarly $Lc_2$ value shifted from 8.2N to 9N and further to 9.5N with similar increase in applied loading rates.

For SS samples, the $Lc_1$ value changed from 15 to 17N and $Lc_2$ value changed from 22N to 24N when the applied loading rate was increased successively from 20N/min to 80N/min.

### 3.5.4 Depth of Penetration

Depth of penetration increased with the increase in applied load. At 30N load, on an average, MS samples had 20-30 µm depth of penetration while SS samples had 12-25 µm depth of penetration. Depth of penetration includes elastic as well as plastic deformation during loading. Besides, there could be error factors such as natural slopes of the samples (thickness variation in sample), mounting errors etc.

### 3.5.5 Effect of Biasing

Increase in biasing voltage from 0 to –75V (in a step of 25V), keeping other factors constant during deposition of Nb-N coatings on SS substrates, led to increase in the values of $Lc_1$ and $Lc_2$. However at -100V, the coating became brittle and $Lc_1$ and $Lc_2$ both values dropped drastically. The values are shown in Table 5.

### 3.5.6 Effect of $N_2$ Flow

Nb-N coatings on SS samples with increase in $N_2$ (to Ar) flow from 10% to 70%, keeping the biasing voltage constant at -50V, were tested for scratch adhesion testing. Results with respect to critical loads are shown in Table 6. Coatings deposited at 20 and 30% $N_2$ flow show better adhesion with higher critical loads.

### 3.6 Corrosion Resistance

It is difficult to deposit the hard coatings by physical vapour deposition (PVD) techniques without any porosity. Thus, when a PVD coated sample is exposed to the corrosive environment, the electrochemical behavior of the coated sample is the combined behavior of the coating and the substrate. The polarization curve of such a specimen may be considered as a combination of two curves - one representing the base material, and the other the coating.

### 3.6.1 Open Circuit Potential

Figure 14 shows the changes in open circuit potential (OCP) with immersion time for NbN and NbN with Nb interlayer coated on MS sample. For NbN coating potential decreased from an initial value of -252mV to -478mV in 30min and to -498 mV in 40min after which it remained nearly constant indicating the establishment of equilibrium between metal and the solution. The equilibrium value was found to be quite similar to mild steel substrate without coating, thus indicating the presence of pores in the coating, resulting in corrosion taking place beneath NbN coating. For NbN coatings on MS substrates with Nb interlayer, the OCP shifted to -154mV in the beginning from -252mV for coating without interlayer. The value decreased to about -443mV after 40min and remained constant thereafter.

### 3.6.2 Potentiodynamic tests

Corrosion potential ($E_{corr}$) and corrosion current density ($I_{corr}$), were determined from their polarization curves using the Tafel slopes $\beta_a$ and $\beta_c$ and polarization resistance ($R_p$). A high $E_{corr}$ and a low $I_{corr}$ values are indicative of good corrosion resistance.

Table 7 lists the $E_{corr}$ and $I_{corr}$ values for MS substrate; NbN, Nb and NbN coating with Nb interlayer on MS substrate. Plain MS substrate had $E_{corr}$ value of -496.4mV and $I_{corr}$ value of 1440 $\mu A/cm^2$. Coating the mild steel sample with NbN improved the corrosion resistance by decreasing the corrosion current ($I_{corr}$) to 150 $\mu A/cm^2$ and increasing the $E_{corr}$ (less negative) to -412mV. However, the improvement was constrained due to the presence of pin-hole porosity inherent in PVD coatings. Due to the presence of pin-hole defects in the NbN coatings, rapid pitting corrosion of mild steel substrate takes place at these defects; even leading to partial debonding of the coating during the tests. The potentiodynamic curves for NbN coatings, therefore, mimic the behavior of MS substrate.

Nb interlayer was found to improve the corrosion resistance of NbN coated MS substrates effectively. For Nb interlayered NbN coating $E_{corr}$ and $I_{corr}$ were found to be -396.2mV and 14.8$\mu A/cm^2$ respectively.

With the duplex coating, the adherence also improved because duplex coating did not delaminate for the full potential sweep of -1000mV to +1000mV during the potentiodynamic corrosion test. Potentiodnamic curves for substrate, NbN coating and NbN coating with Nb interlayer on MS substrates are shown in figure 15. Potentiodynamic curve for the duplex coating shows a remarkable improvement in corrosion improvement for MS substrate. In general, it was observed that the duplex coating is superior to NbN coating alone.

## 4. Conclusions

NbN coatings were deposited on SS substrates by reactive DC magnetron sputtering. $N_2$/Ar flow was varied from 0-70% and substrate biasing from zero to -150V in a step of 25V. Coatings were characterized for their thickness by weight gain and calotest method, hardness by Knoop micro hardness tester, phase analysis by X-ray diffraction technique and adhesion by scratch adhesion tester. The effect of $N_2$ flow and substrate biasing were evaluated. After the optimization of process parameters, NbN coating was deposited on MS substrate as top coat with Nb as interlayer deposited by sputtering in 2 μm thickness. Effect of interlayer on NbN coating on MS substrate was studied for improvement in surface hardness by Knoop micro indentation and corrosion by potentiodynamic polarization technique in 1N $H_2SO_4$ solution at room temperature. Surface hardness increased from 1084 $HK_{25}$ for NbN coating on MS to 1618 $HK_{25}$ when Nb as interlayer was incorporated. It was observed that NbN coating alone could not protect the MS substrate effectively. With duplex coating the corrosion resistance increased remarkably. $I_{corr}$ was found to decrease from 150.2μA/cm$^2$ for NbN coating on MS to 14.8μA/cm$^2$ when incorporated with Nb interlayer.

**Table 1 : Deposition parameters for Nb and NbN coatings**

| Parameter | Value |
|---|---|
| Base pressure | $2 \times 10^{-6}$ mbar |
| Operating pressure | $5 \times 10^{-3}$ mbar |
| Argon gas flow rate | 20 sccm |
| Nitrogen gas flow rate | 0–14 sccm |
| Substrate biasing | 0 to -150 V |
| Target current NbN/Nb | 0.25/0.3 Ampere |
| Target-substrate distance | ~ 60 mm |

**Table 2 : Hardness ($HK_{25}$) and depth of indentation for NbN coatings on SS**

| $F_{N2} / F_t$ % | Hardness | Depth (μm) |
|---|---|---|
| 5 | 1450 | 0.522 |
| 10 | 1710 | 0.480 |
| 20 | 2040 | 0.440 |
| 30 | 2028 | 0.441 |
| 40 | 1885 | 0.458 |
| 50 | 1732 | 0.477 |
| 60 | 1628 | 0.492 |
| 70 | 1572 | 0.501 |

**Table 3 : Surface hardness for Nb, NbN and duplex coatings on MS**

| Coating | 100gf | 50gf | 25gf |
|---|---|---|---|
| Substrate | 182 | 186 | 198 |
| Nb | 332 | 362 | 434 |
| NbN | 487 | 674 | 1084 |
| Nb+NbN | 712 | 905 | 1436 |

**Table 4 : μ at different loads for NbN coatings on MS and SS during scratch tests**

| Load(N) | MS | SS |
|---|---|---|
| 20 | 0.23 | 0.20 |
| 30 | 0.28 | 0.25 |
| 40 | 0.35 | 0.30 |
| 60 | 0.45 | 0.40 |

**Table 5 : Effect of biasing on critical loads during scratch tests for NbN coatings on SS**

| Biasing (-V) | $Lc_1$ | $Lc_2$ |
|---|---|---|
| 0 | 6.5 | 10.5 |
| 25 | 10.5 | 20 |
| 50 | 12.0 | 24 |
| 75 | 15.6 | 26 |
| 100 | 7.0 | 11 |

**Table 6 : Effect of $N_2$ flow on critical loads during scratch tests for NbN coatings on SS**

| $N_2$/Ar flow % | $Lc_1$ | $Lc_2$ |
|---|---|---|
| 10 | 11 | 18 |
| 20 | 12 | 24 |
| 30 | 11 | 25 |
| 40 | 10 | 20 |
| 50 | 8 | 20 |
| 60 | 8 | 18 |
| 70 | 7 | 14 |

**Table 7 : $E_{corr}$ and $I_{corr}$ values for single and duplex coatings**

| Coating | $E_{corr}$ (mV) | $I_{corr}$ (µA/cm$^2$) |
|---|---|---|
| Substrate | -496.4 | 1440.3 |
| NbN | -412.1 | 150.2 |
| Nb | -469.3 | 121.8 |
| Nb+NbN | -396.2 | 14.8 |

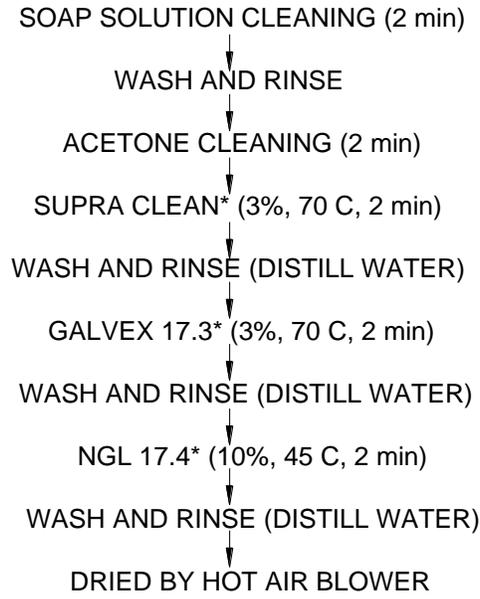

**Fig. 1 : Cleaning cycle adopted**

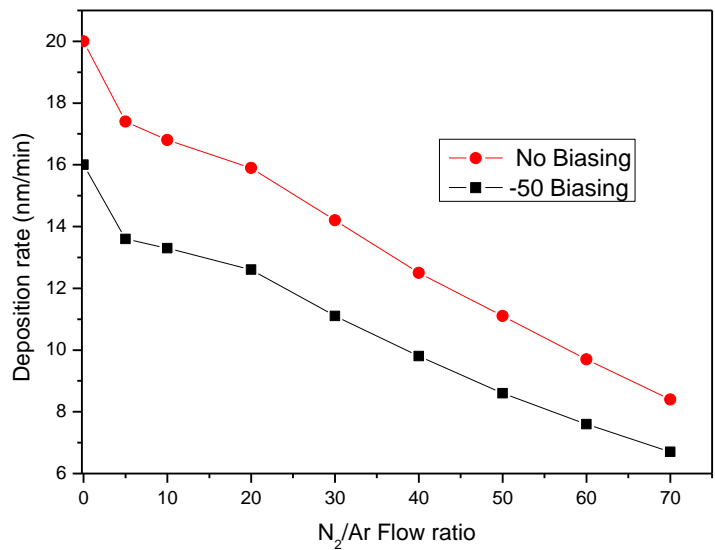

**Fig. 2 : Deposition rate of Nb-N coatings Vs $N_2$ flow**

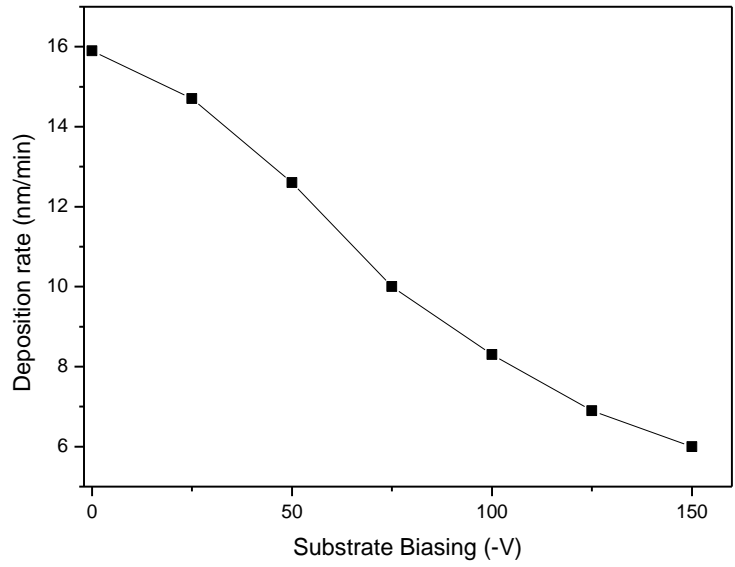

**Fig. 3 : Deposition rate of Nb-N coating Vs substrate biasing**

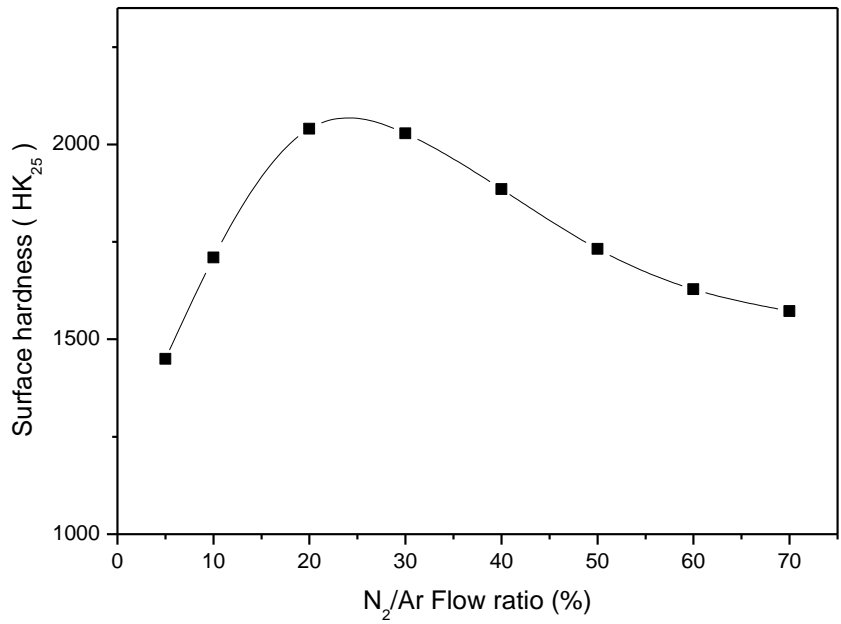

**Fig. 4 : Surface hardness of Nb-N as a function of $N_2$ flow**

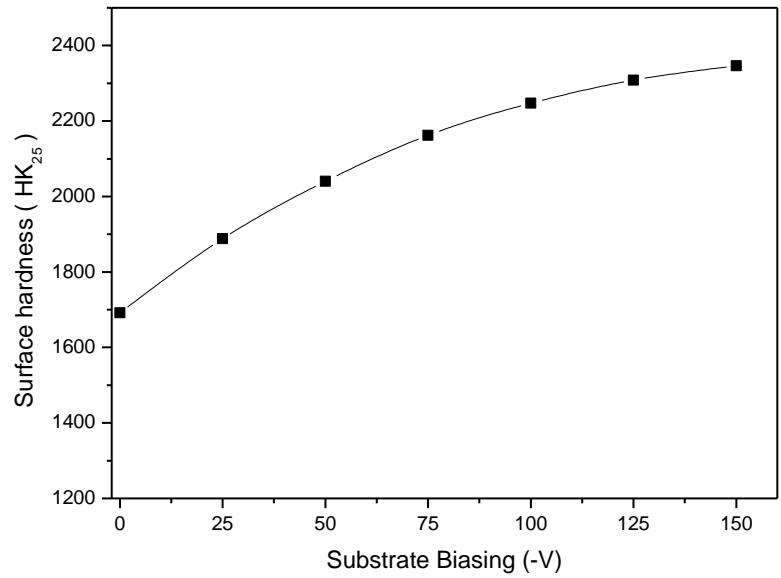

**Fig. 5 : Surface hardness of Nb-N as a function of substrate biasing**

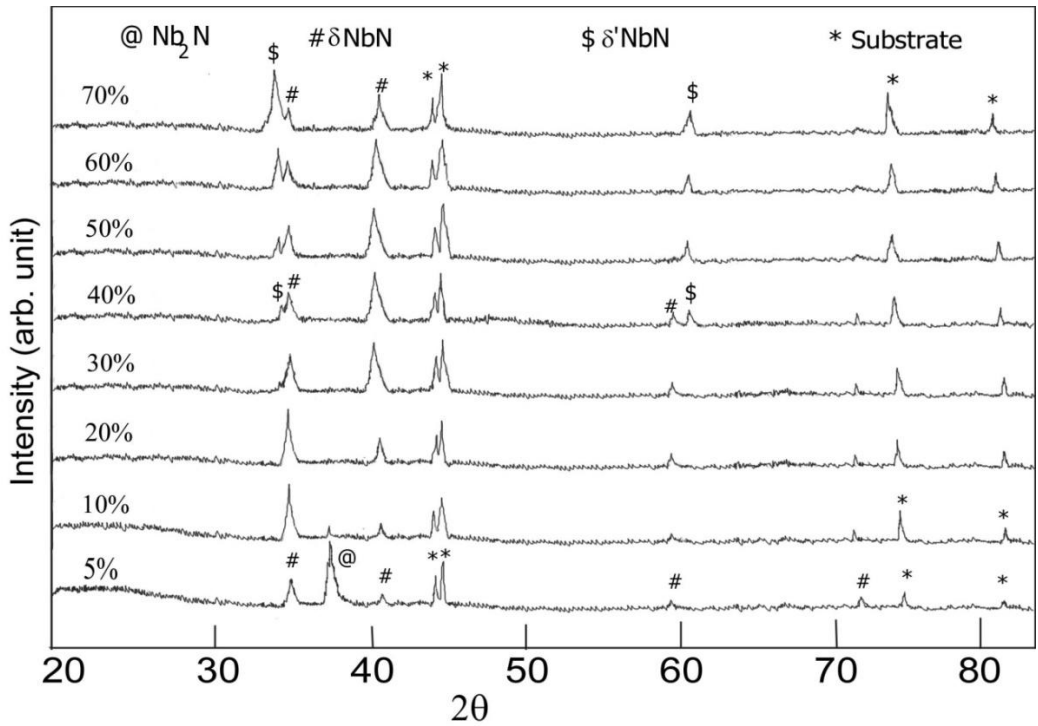

**Fig. 6 : X-ray diffraction patterns of Nb-N Coatings Vs $N_2$ Flow**

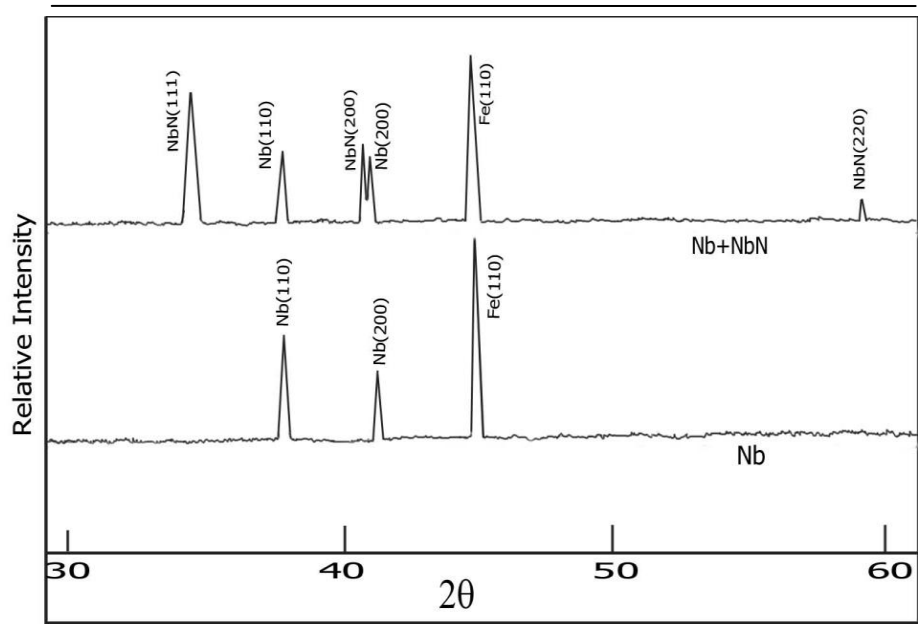

**Fig. 7 : X-ray diffraction patterns for Nb and NbN with Nb interlayer on MS**

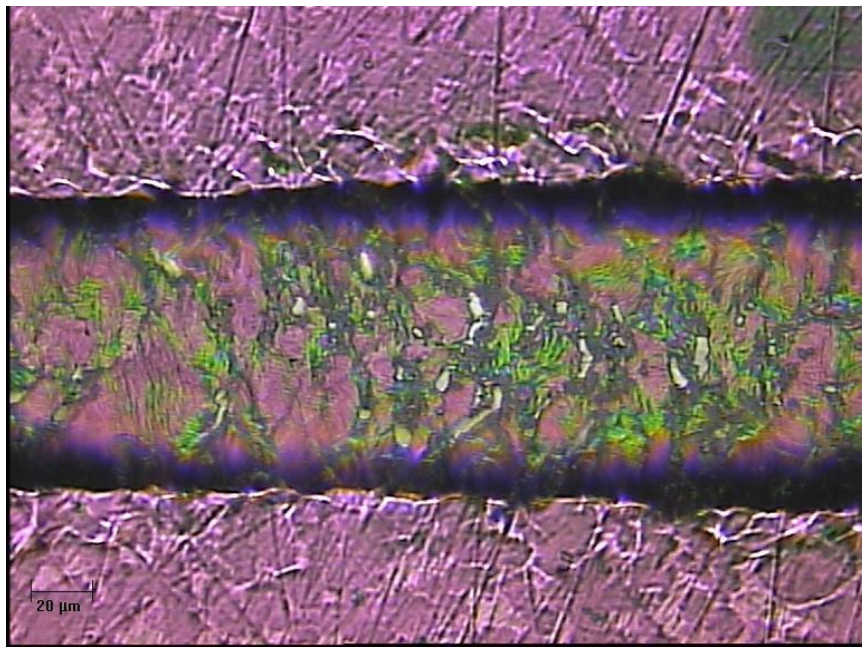

**Fig. 8 : Scratch test for NbN coating on MS showing chipping and cracks (9N)**

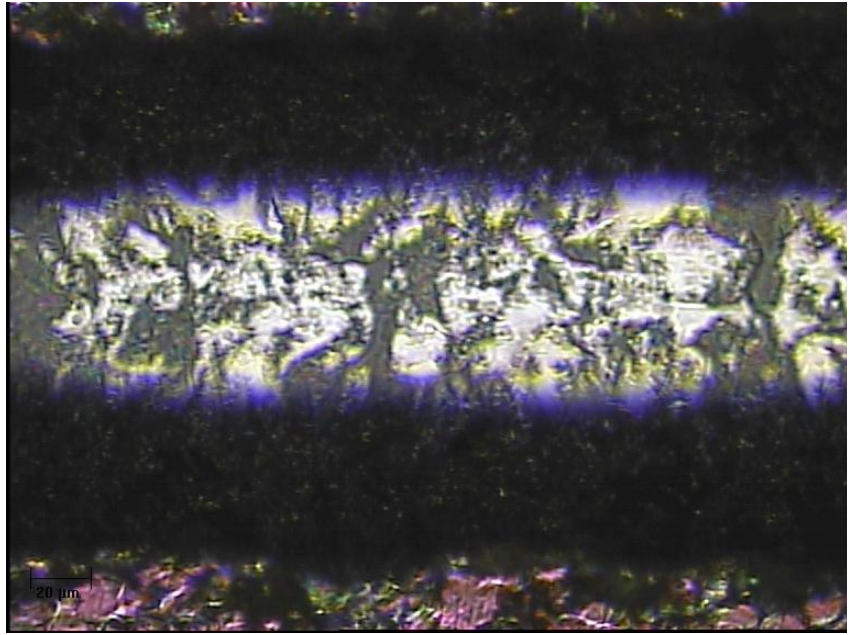

**Fig. 9 : Scratch test for NbN coating on MS showing nearly complete delamination (28N)**

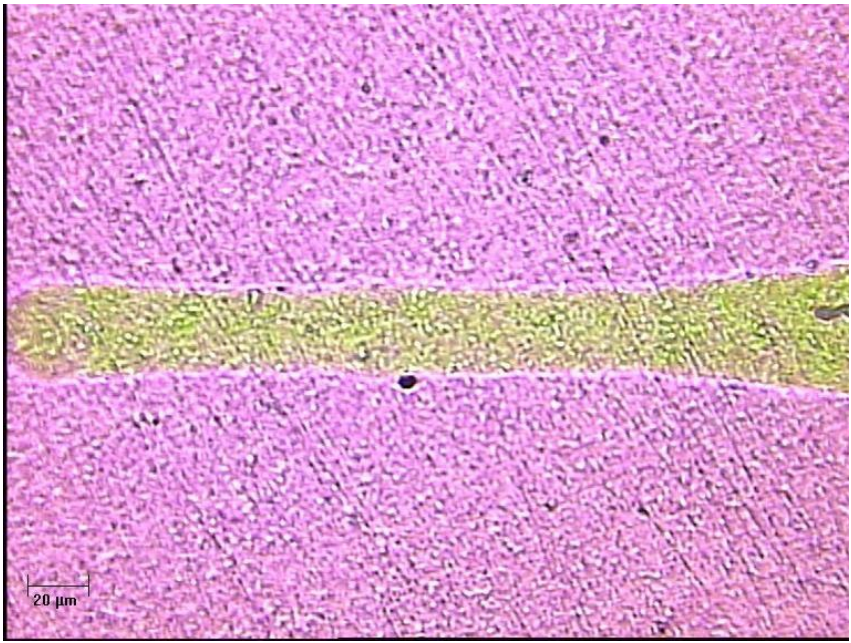

**Fig. 10 : Scratch test for NbN coating on SS showing start of the scratch (1N)**

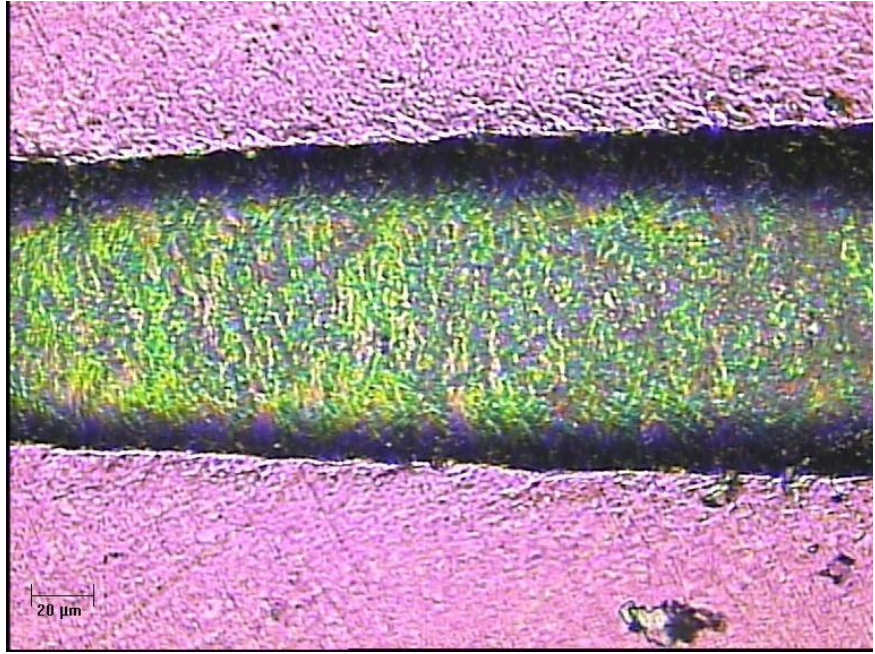

**Fig. 11 : Scratch test for NbN on SS revealing segregation, cracks, pores (12N)**

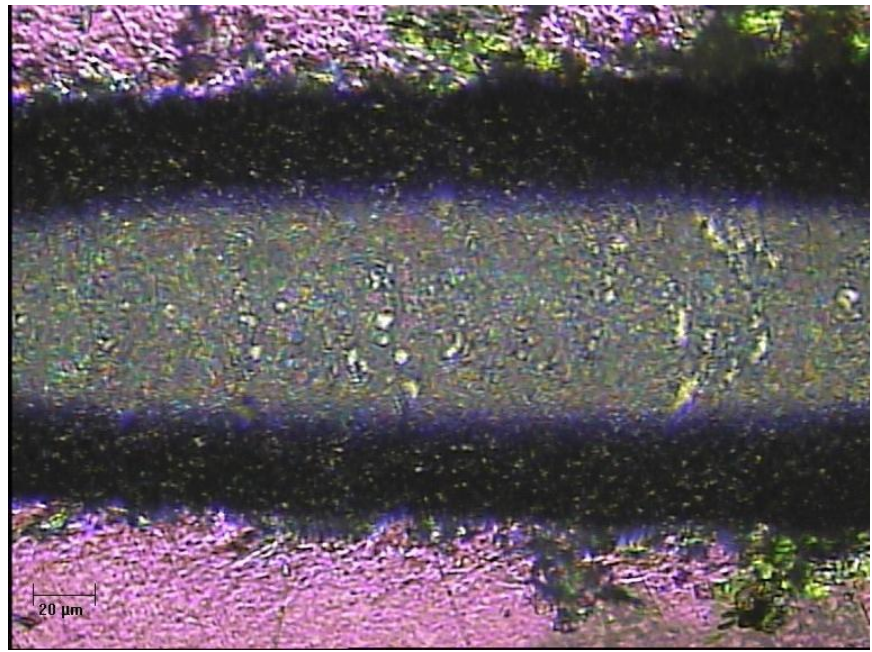

**Fig. 12 : Scratch test for NbN on SS - chipping, cracks, pores, pile-up (22N)**

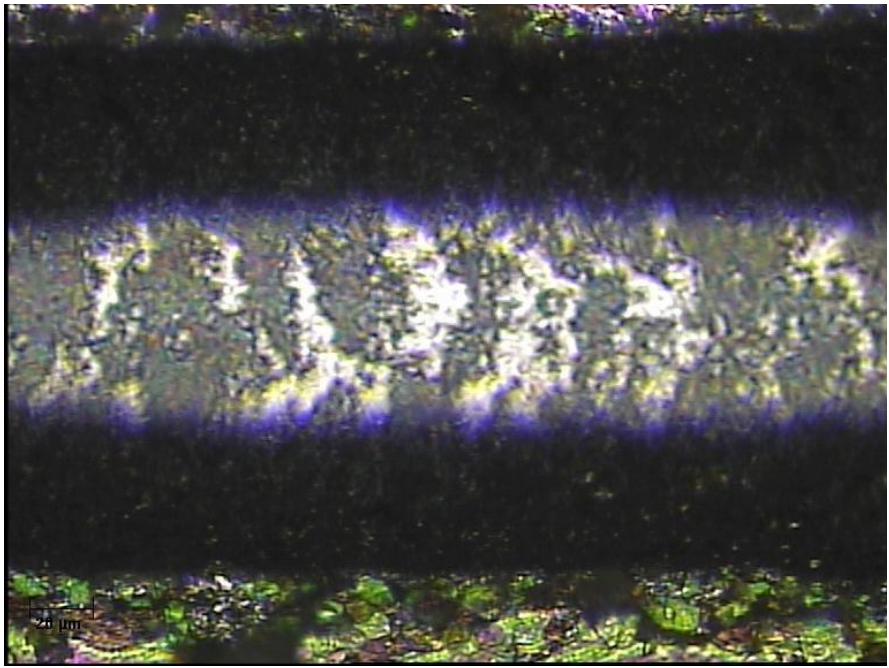

**Fig. 13 : Scratch test for NbN on SS – chipping, cracks, pores, pile-up, delamination (32N)**

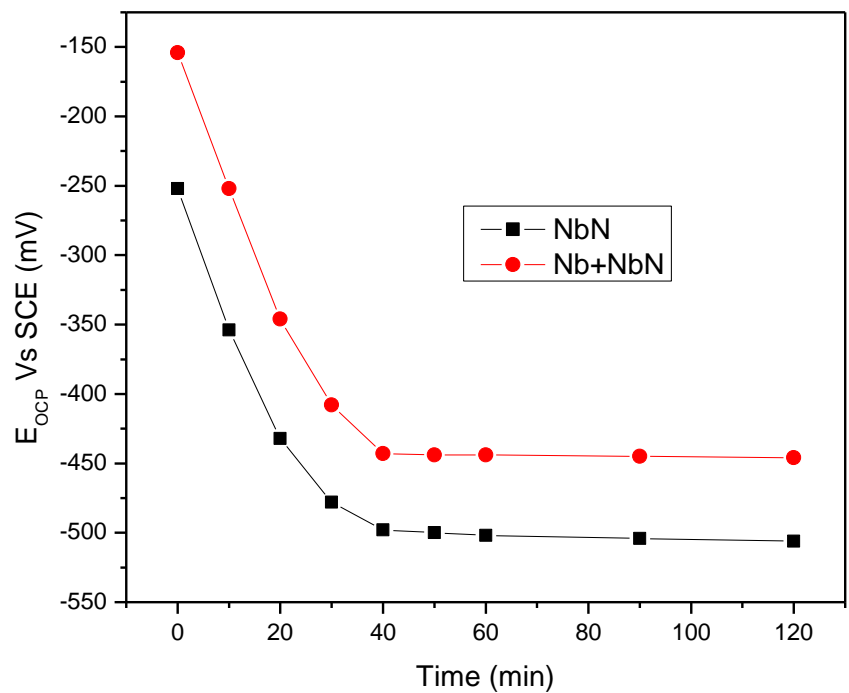

**Fig. 14 : $E_{OCP}$ Vs time for NbN and NbN with Nb interlayer on MS samples in 1N $H_2SO_4$.**

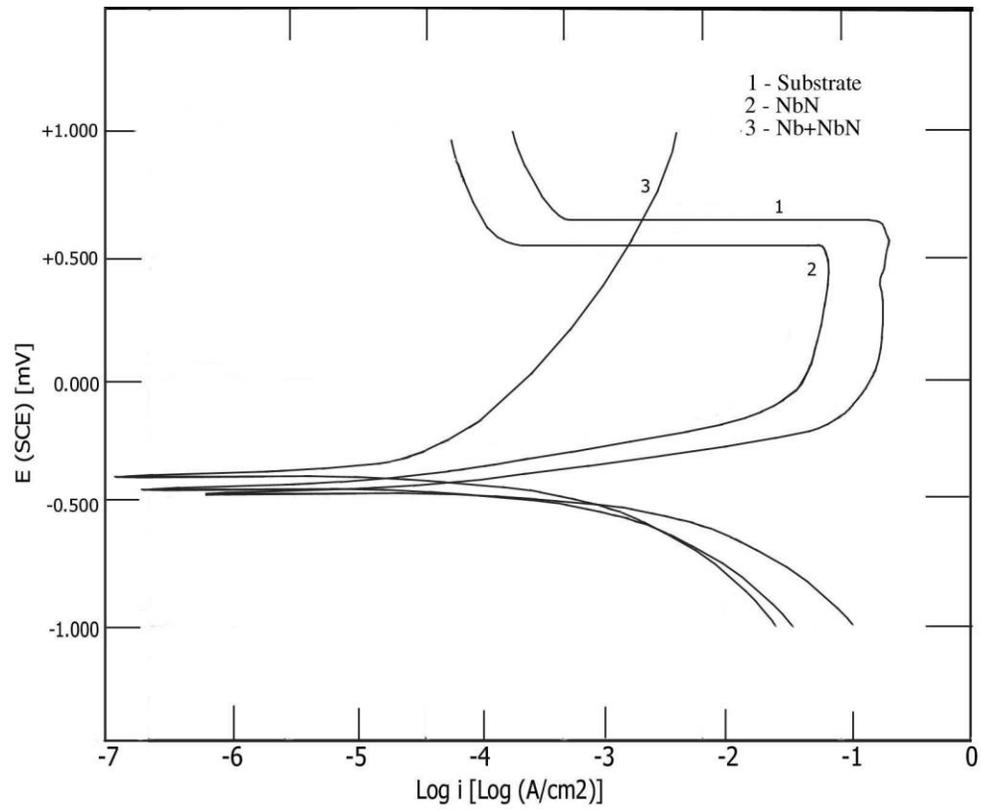

**Fig. 15 : Potentiodynamic curves for MS, NbN and NbN with Nb interlayer on MS**